\documentclass{mn2e}
\usepackage[utf8x]{inputenc}
\usepackage[english]{babel}
\usepackage{times}
\usepackage[T1]{fontenc}
\usepackage{array}
\usepackage{amsmath}
\usepackage{graphicx}

\title[Magnetic field evolution]{Magnetic field evolution of accreting neutron stars}
\author[Ya.~N. Istomin, I.~A. Semerikov]{Ya.~N. Istomin $^{1,2,3}$, I.~A. Semerikov $^{2}$ \\
${1}$ P.N.~Lebedev Physical Institute, Leninsky Prospect 53, Moscow 119991, Russia \\
${2}$ Moscow Institute Physics and Technology, Institutskii per. 9, Dolgoprudnyi, Moscow region, 141700, Russia \\
${3}$ E-mail: istomin@lpi.ru \\}

\begin{document}
\date{}
\pagerange{\pageref{firstpage}--\pageref{lastpage}}
\pubyear{2015}
\maketitle
\label{firstpage}

\begin{abstract}

The flow of a matter, accreting onto a magnetized neutron star, is
accompanied by an electric current. The closing of the electric
current occurs in the crust of a neutron stars in the polar region
across the magnetic field. But the conductivity of the crust along
the magnetic field greatly exceeds the conductivity across the
field, so the current penetrates deep into the crust down up to
the super conducting core. The magnetic field, generated by the
accretion current, increases greatly with the depth of penetration
due to the Hall conductivity of the crust is also much larger than
the transverse conductivity. As a result, the current begins to
flow mainly in the toroidal direction, creating a strong
longitudinal magnetic field, far exceeding an initial dipole
field. This field exists only in the narrow polar tube of $r$
width, narrowing with the depth, i.e. with increasing of the crust
density $\rho$, $r\propto \rho^{-1/4}$. Accordingly, the magnetic
field $B$ in the tube increases with the depth, $B\propto
\rho^{1/2}$, and reaches the value of about $10^{17}$ Gauss in the
core. It destroys super conducting vortices in the core of a star
in the narrow region of the size of the order of ten centimeters.
Because of generated density gradient of vortices they constantly
flow into this dead zone and the number of vortices decreases, the
magnetic field of a star decreases as well. The attenuation of the
magnetic field is exponential, $B=B_0(1+t/\tau)^{-1}$. The
characteristic time of decreasing of the magnetic field $\tau$ is
equal to $\tau\simeq 10^3$ years. Thus, the magnetic field of
accreted neutron stars decreases to values of $10^8 - 10^9$ Gauss
during $10^7-10^6$ years.

\end{abstract}



\section{Introduction}

Rather quickly after the discovery of radio pulsars (Hewish et al. 1968),
it became clear that their sources are neutron stars, and the
mechanism of radiation is directly associated with rotating
magnetic field frozen into the body of a star (Goldreich \& Julian 1969).
Magnetic fields of neutron stars: radio pulsars, x-ray pulsars and
x-ray variables, are measured mainly by two methods. The first
method is based on the assumption that a neutron star with a
frozen-in magnetic field is an oblique magnetic rotator. Then the
loss of the energy by the magneto dipole radiation equals
$$
{\dot E}=-\frac{B_0^2\Omega^4 R^6}{6c^3}\sin^2\chi.
$$
Here $R$ is the radius of the star ($R\simeq 10 km$), $\chi$ is the angle between the rotation axis and the axis of the magnetic dipole,
$B_0$ is the magnetic field at the surface on the magnetic pole, $\Omega$ is the frequency of the star rotation, $c$ is the speed of the light.
Considering that the energy source is the rotation of the star, and taking into account that
$E=I\Omega^2/2$, where $I$ is the moment of inertia of the star ($I\simeq 10^{45} g cm^2$),
measuring the rotation period $P=2\pi/\Omega$ and period derivative ${\dot P}$, we obtain the estimation for the magnitude
of the magnetic field $B_0$,
\begin{equation}\label{estmateB}
B_0=\left(\frac{3}{2\pi^2}\frac{Ic^3}{R^6}P{\dot P}\right)^{1/2}\simeq 10^{12}\left(\frac{P}{1s}\right)^{1/2}
\left(\frac{{\dot P}}{10^{-15}}\right)^{1/2} G.
\end{equation}
In this expression it is considered that the angle of the
inclination of the magnetic dipole $\chi$ is of the order of
unity. The above evaluation (\ref{estmateB}) suggests that the
expression for the magneto dipole losses of a dipole rotating in a
vacuum is also true for the neutron star with a magnetosphere
filled by a dense electron-positron plasma, which is born in a
strong magnetic field. As was shown by Beskin, Gurevich \& Istomin (1993),
the electromagnetic radiation is screened
by the magnetospheric plasma, and the loss of rotation are
determined by the electric currents flowing in the magnetosphere
and on the stellar surface. However, the relation (\ref{estmateB})
remains in force for magnetospheric electric currents of the order
of the Goldreich-Julian current, $j_{GJ}=B\Omega/2\pi$ (see
the paper by Beskin, Istomin \& Philippov 2013). The second method is to measure the
absorption lines in the spectrum of x-ray pulsars. It is assumed
that they are formed by cyclotron absorption in the atmosphere of
neutron stars when the wave frequency $\omega$ coincides with the
$n$ harmonic of the cyclotron frequency of cold electrons,
$\omega=neB/m_e c$. Here $e$ and $m_e$ is the electric charge and
the mass of electron, respectively.

\begin{figure}
\begin{center}
\includegraphics[width=8cm]{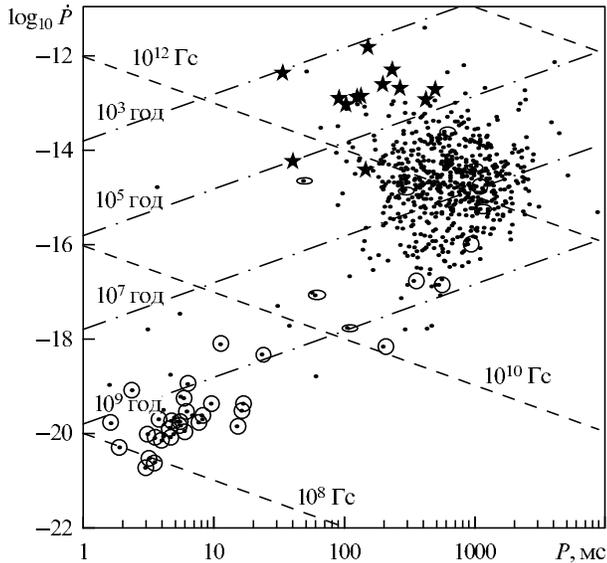}
\end{center}
\caption{The diagram of distribution of radio pulsars in the plane $P$ - ${\dot P}$ (the period-the period derivative). Points circled by
circles and ellipses correspond to sources in
binary systems. Pulsars associated with supernova remnants marked by asterisks (Lorimer 2001).
The dashed lines correspond to different values of the magnetic field and the age $P/2{\dot P}$.}
\label{fig1}
\end{figure}

Measured by these methods, magnetic fields of pulsars are in the
range of $10^{8} - 10^{13}$ Gauss, see Fig. \ref{fig1}
(Lorimer, 2001). One can distinguish two groups of neutron stars -
single radio pulsars with fields $10^{11} - 10^{13}$ Gauss and
recycled pulsars with fields of $10^8 - 10^{10}$ Gauss, which were
or are members of close binary systems. We are not considering
here a separate group of neutron stars with super strong magnetic
fields, $10^{13} - 10^{15}$ Gauss, so-called magnetars, that emit
energy stored in the magnetic field, far exceeding the
rotational energy.

It is seen that the neutron star with a low magnetic field, $10^8 - 10^{10}$ Gauss, do have small values of the braking,
${\dot P}\simeq 10^{-19} s/s$, and hence large dynamic life times, $\tau_d=P/2{\dot P}\simeq 10^{9}$ years, despite
they have small periods of rotation (the majority of them are millisecond pulsars). Neutron stars with the traditional magnetic
fields, $10^{11}-10^{13}$ Gauss, have a life time of $\tau_d\simeq 10^7$ years. This time for them is almost identical with
the so-called
kinematic life time, $\tau_k=z/v$, where the value of $z$ is the height of the neutron star above the Galactic plane,
and $v$ is their proper velocity, normal to the plane of the Galaxy. Pulsars are fast objects with velocities
$\simeq 200-400$ km/s (Lyne, Anderson \& Salter 1982), so their normal speeds are of the order of $\simeq 100$ km/s. From the fact of correlation
between times $\tau_d$ and $\tau_k$, it follows that the time of evolution (decay) of the magnetic field of neutron stars $\tau_B$ is of the
value of $\tau_B>10^7$ years.
Based on these properties, we can estimate the lower limit of the
conductivity of the matter of a neutron star $\sigma$. Magnetic
viscosity $\eta=c^2/4\pi\sigma$ must be of the order of
$R^2/\tau_B$, i.e. $\eta<3\cdot 10^{-3} cm^2/s$. This gives the
estimate of $\sigma>3\cdot 10^{22} s^{-1}$. The conductivity of
the crust of neutron stars is about $10^{20} s^{-1}$ (Blandford,
Appelgate \& Hernquist 1983), which means
that electric currents, which create the magnetic field $B$ frozen
into the body of the star, do not flow in the crust. Moreover,
since the crust thickness $H$ is about $1 km\simeq R/10$, then the
condition for the crust conductivity becomes even more hard,
$\sigma>3\cdot 10^{24} s^{-1}$. All these mean that the source of
the magnetic field of neutron stars is placed in the core, where
the matter is in the super conducting state. Nonetheless, it is
clear that neutron stars, passed or are in the process of
accretion, lost a significant part of their magnetic field. And
this phenomenon cannot be due to processes on the surface and in
outer layers of the crust (motion of matter, heating, etc.)
associated with accretion of the matter. The magnetic field,
originating in the core of neutron stars, cannot be screened by
the surface currents due to the continuity of the radial component
of the magnetic field. How does accretion affect the core of the
star? The only possibility is the effect of additional
electromagnetic fields arising from the accretion of the matter
onto the surface of the neutron star.

It should be noted that accretion of a conducting matter (plasma) onto a
magnetized star is very different from accretion of a neutral gas
or accretion in the absence of a magnetic field in the magnetosphere
of a star. Conducting accretion disk closes different magnetic surfaces
of the magnetic field of a star. The magnetic field lines, emerging from
a stellar surface closer to the magnetic pole, come to more distant parts
of a disk from a star. While more equatorial lines come to the inner part
of a disk. The magnetic fluxes $f$ inside these magnetic surfaces are
different, their difference is equal to $\Delta f$. Then due to the rotation of
magnetic field lines together with a star, there appears
the difference of the potential of the electric field, $\psi=\Omega\Delta f/c$.
That is the so-called unipolar inductor (Landau, Lifshitz, Pitaevskii 1984). The electrical
voltage $\psi$ produces the electric current $j$ flowing in the magnetosphere
along magnetic field lines due to strong particle magnetization. The closer
of this current occurs on a disk in the outer region, and on a star in its polar cap.
The system 'disk-rotating star with frozen-in magnetic field' forms a simple dynamo
machine, in which, depending on the direction of the electric current, a disc spins
a star or a star spins a disk (propeller). The presence of electric current does
not mean appearance of electric charge in plasma, i.e. charge separation. Charge
density and electric current are independent characteristics. It is clear that
any electric charge will cause the appearance of the electric field, which will
restore the quasi-neutrality of a plasma. The stellar magnetosphere has a 'sea' of
electrons which neutralize any charge.

The magnetic field in a magnetosphere fundamentally changes the motion
of charged particles. In a magnetic field the generalized angular momentum of a
particle consists of two parts: the mechanical angular momentum $l$
and the electromagnetic one $qf/c$. It is similar to the particle generalized
momentum is the sum of the mechanical momentum $p$ and the electromagnetic part
$qA/c$. Here q is the particle charge and A is the vector potential of the
electromagnetic field. But electromagnetic part of the angular momentum in the
magnetosphere near a stars far exceeds mechanical angular momentum, their ratio
is of the order of ratio of the cyclotron frequency of particle rotation in magnetic
field $\omega_c$ to the frequency of rotation of a star $\Omega$, $\omega_c/\Omega>>1$.
This means that a charged particle, having a mechanical angular momentum far
away from the star, transforms it into electromagnetic one near a star. And transmission
of electromagnetic angular momentum from the plasma of a disk to a star (and vice versa)
means existence of the flux of the moment of the Pointing vector, which does not arise
without generation of additional to the stellar magnetic field electromagnetic fields
and electric currents.

\begin{figure}
\begin{center}
\includegraphics[width=8cm]{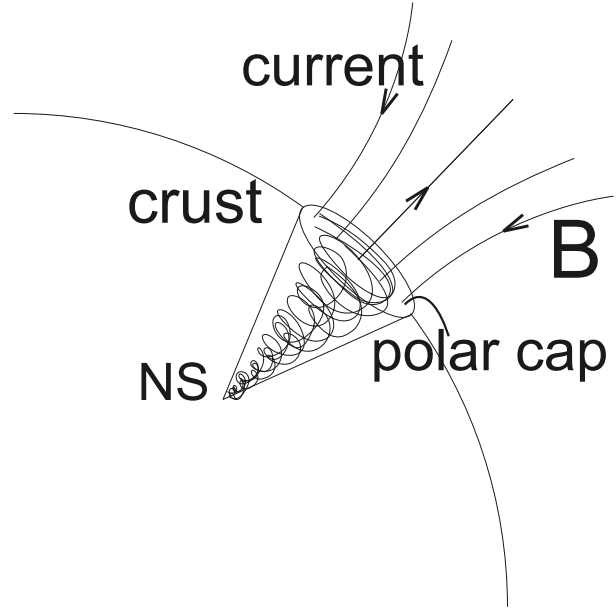}
\end{center}
\caption{Magnetic field and electric current in the polar region.}
\label{fig2}
\end{figure}

The accretion of the matter onto the surface of a magnetized star
occurs in the region of the magnetic poles of the size of
$r_0\simeq 10^3$ (see Fig. \ref{fig2}). As we explained above, the accretion is accompanied
by the electric current,  This electric current
forms a closed loop, it flows through the accretion disk in the direction to the star, then in the
magnetosphere along the magnetic surfaces, which
link the internal edge of the disk with the polar cap on the surface, then returns to the
disk at the inner accretion flow for the case of a corotating disk.  This current transports the angular
momentum from the accretion disk to the star. Its closing in the polar
cap across the magnetic field produces the Ampere force spinning up
(for a corotating disk) or decelerating (for a counterrotating disc) the stellar
rotation. The torque $K$, acting on the star, equals
\begin{equation}\label{StarTorque}
K={\dot M}r_d^2\Omega.
\end{equation}
Here ${\dot M}$ is the accretion rate and the value of $r_d$ is the inner radius of the accretion disk,
which for corotation is equal to the radius of corotation
$r_c$,
$$
r_c=\left(\frac{GM}{\Omega^2}\right)^{1/3}=1.5\cdot 10^8\left(\frac{M}{M_\odot}\right)^{1/3}\left(\frac{P}{1s}\right)^{2/3} \, cm,
$$
and for counterrotation equals the Alfven radius $r_A$,
\begin{eqnarray}
r_A=\left(\frac{B_0^2 R^6}{2^{1/2}{\dot M}r_c^{3/2}\Omega}\right)^{2/7}=1.6\cdot 10^9\left(\frac{B_0}{10^{12}G}\right)^{4/7} \cdot \\  \nonumber
\left(\frac{{\dot M}}{10^{-10} \, M_\odot/yr}\right)^{-2/7}\left(\frac{r_c}{10^8 cm}\right)^{-3/7}\left(\frac{P}{1s}\right)^{2/7} \, cm. \nonumber
\end{eqnarray}
The value of $G$ is the gravitational constant, $M$ is the mass of the star.
The physics of the plasma accretion onto a magnetized star is described in details by Istomin \& Haensel (2013).

Closing of the current occurs in the crust of neutron stars because
the conductivity of the plasma in the accretion column, $\sigma_p=3\cdot 10^{17}(T_e/1KeV)^{3/2}(\Lambda/10)^{-1}\,s^{-1}$ ($T_e$
is the electron temperature,
$\Lambda$ is the Coulomb logarithm, $\Lambda\simeq 20$), significantly less than the conductivity of the crust. But the conductivity of
the crust is strongly anisotropic, the conductivity
along the magnetic field, $\sigma_\parallel$, is significantly higher than the conductivity across the magnetic field,
$\sigma_\perp$, $\sigma_\parallel>>\sigma_\perp$.
Therefore, the accretion current must penetrate deep into the crust, the depth of the penetration must be of the order of
$(\sigma_\parallel/\sigma_\perp)r_0>>r_0$, $r_0$ is the radius of the polar cap.
In addition, due to the strong magnetization of electrons, the crust also has the Hall conductivity, $\sigma_\wedge>>\sigma_\perp$, and
the electric current begins to flow mainly in the azimuthal direction, creating the magnetic field directed along
the axis of the penetration, i.e. along the axis of the dipole (Fig. \ref{fig2}). Thus, the electric current $I$, which is
necessary to create the magnetic field $B_0$ in the polar region, equals

$$
I=\frac{B_0 r_0 c}{2}\left(\frac{\sigma_\wedge}{\sigma_\perp}\right)^{-1}.
$$
It creates the torque $K$,
\begin{equation}\label{Torque}
K=\frac{IB_0r_0^2}{2}.
\end{equation}
Equating the values of (\ref{StarTorque}) and (\ref{Torque}), we obtain
\begin{eqnarray}\label{Current}
I=\left[\frac{\pi}{2}\frac{{\dot M}B_0r_d^2c^3}{P}\left(\frac{\sigma_\wedge}{\sigma_\perp}\right)^{-2}\right]^{1/3}= \\ \nonumber
3.2\cdot 10^{14}
\left(\frac{{\dot M}}{10^{-10} \, M_\odot/yr}\right)^{1/3} \cdot \\ \nonumber
\left(\frac{B_0}{10^{12}G}\right)^{1/3}\left(\frac{r_d}{10^8 cm}\right)^{2/3}
\left(\frac{P}{1s}\right)^{-1/3}\left(\frac{\sigma_\wedge}{30\sigma_\perp}\right)^{-2/3} \, A.
\end{eqnarray}
The size of the polar cap equals
\begin{eqnarray}\label{PolarCap}
r_0=\left[\frac{4\pi{\dot M}r_d^2}{B_0^2 P}\left(\frac{\sigma_\wedge}{\sigma_\perp}\right)\right]^{1/3}=1.3\cdot 10^3
\left(\frac{{\dot M}}{10^{-10} \, M_\odot/yr}\right)^{1/3} \cdot \\  \nonumber
\left(\frac{B_0}{10^{12}G}\right)^{-2/3}\left(\frac{r_d}{10^8 cm}\right)^{2/3}
\left(\frac{P}{1s}\right)^{-1/3}\left(\frac{\sigma_\wedge}{30\sigma_\perp}\right)^{1/3} \, cm.
\end{eqnarray}
The transverse size $r$ of the region, where the current $I$ is closed, becomes less with the distance from the stellar surface while the
current penetrates into the crust.
At the same time, the magnetic field, which equals $B_0$ on the surface, increases with depth, $B=B_0(r_0/r)^2$. But it exists only in the
narrow tube, not
penetrating into other areas of the crust. Such a structure is similar to a needle and can be called the 'magnetic needle' (Fig. \ref{fig2}).

\section {Closing of the current in the crust}

In this section we will describe in details how the accretion current is closed in the crust in the polar region. As we mentioned above,
the conductivity of the crust along the stellar magnetic field, which is almost perpendicular to the stellar surface in polar cap,
is much larger than the crust conductivity across the magnetic field. It means that the current must penetrate deep to the crust.
The polar cap on the surface of the star, onto which a plasma accretes from a disk, has high degree of the azimuthal symmetry.
Therefore, we consider the axisymmetric problem of closing of currents in the crust in the region near the axis of the magnetic dipole.
In this case a closed electric current, $\nabla {\bf j} = 0$, is presented in the form
\begin{equation}\label{CurDens}
{\bf j}=\nabla J_1 \times \nabla \varphi + J_2 \nabla \varphi,
\end{equation}
where $J_1$ and $J_2$ are scalar functions of the depth $h$ and the distance to the axis of the dipole $r$, $J_1=J_1(h,r); \, J_2=J_2(h,r)$,
and $\varphi$ is the azimuthal angle. Since $\nabla{\bf B}=0$, the magnetic field ${\bf B}$ is also representable in
the form
\begin{equation}
{\bf B}=\nabla f \times \nabla \varphi + g \nabla \varphi,
\end{equation}
where $f(h,r)$ is the poloidal flux of the magnetic field, and $g(h,r)/r$ is the toroidal magnetic field.
From the Maxwell's equation $curl{\bf B}=4\pi{\bf j}/c$ we get
\begin{equation}\label{Maxwell}
J_2=-\frac{c}{4\pi} \widetilde\Delta f; \,  J_1=\frac{c}{4\pi}g, \,   \widetilde\Delta = r \frac{\partial}{\partial r}\left(\frac{1}{r}
\frac{\partial}{\partial r}\right)+\frac{\partial^2}{\partial h^2}.
\end{equation}
The Ohm's law for an anisotropic medium is
\begin{equation}\label{Ohm}
{\bf j}=-{\bf b}(\sigma_{\Vert}-\sigma_{\bot})({\bf b}\nabla \Psi)-\sigma_{\bot}{\nabla \Psi}-\sigma_{\wedge}(\nabla\Psi\times{\bf b}),
\end{equation}
where the vector ${\bf b}$ is the unit vector along the magnetic field,
$$
{\bf b}=r \frac{\nabla f \times \nabla \varphi + g \nabla \varphi}{(\nabla f^2 + g^2)^{1/2}}.
$$
The value of $\Psi(h,r)$ is the potential of the electric field,
which occur in the crust when the electric current of accretion
invades into it. The conductivity along the magnetic field is
$\sigma_{\Vert}$, the transverse conductivity is $\sigma_{\bot}$
and the Hall conductivity is $\sigma_{\wedge}$. The toroidal
component of the equation (\ref{Ohm}) together with equations
(\ref{CurDens}, \ref{Maxwell}) gives the equation
\begin{equation}\label{base}
\frac{c}{4\pi}\widetilde\Delta f=\frac{r g(\sigma_{\Vert}-\sigma_{\bot})}{{\nabla f^2 + g^2}}(\nabla \Psi \times \nabla f)_{\varphi}-
\frac{r \sigma_{\wedge}}{(\nabla f^2 + g^2)^{1/2}}\nabla \Psi \cdot \nabla f,
\end{equation}
while the poloidal components lead to the equation
\begin{eqnarray}\label{PoloidalEq}
\frac{c}{4\pi} \nabla g \times \nabla \varphi =-\frac{r (\sigma_{\Vert}-\sigma_{\bot})(\nabla \Psi \times \nabla f )_{\varphi}}{\nabla f^2 +
g^2}\nabla f \times \nabla \varphi -  \\  \nonumber
\sigma_{\bot}\nabla\Psi-\frac{r g \sigma_{\wedge}}{(\nabla f^2 + g^2)^{1/2}} \nabla \Psi \times
\nabla \varphi.
\end{eqnarray}
Let us find the value of $(\nabla \Psi \times \nabla f)_{\varphi}$ from the equation (\ref{base})
\begin{equation}\label{PhiComp}
(\nabla \Psi \times \nabla f)_{\varphi}=\frac{\nabla f^2 + g^2}{r g (\sigma_{\Vert}-\sigma_{\bot})}\left[ \frac{c}{4\pi}\widetilde\Delta f +
\frac{r \sigma_{\wedge}}{(\nabla f^2 + g^2)^{1/2}}\nabla \Psi \cdot \nabla f \right].
\end{equation}
The conductivity along the magnetic field lines is large, $r\sigma_{\Vert}/c \gg 1 \, (\sigma_{\Vert}>>\sigma_{\bot})$,
i.e. $(\nabla \Psi \times \nabla f)_{\varphi}\simeq 0.$
This means that the electric potential $\Psi$ is constant along the magnetic surface $f=const, \Psi=\Psi(f)$. Thus, the electric field
is directed perpendicular to the poloidal magnetic field, $\nabla\Psi=(d\Psi/df)\nabla f$.
Substituting (\ref{PhiComp}) into (\ref{PoloidalEq}), we obtain (Istomin, Smirnov \& Pak 2005)
\begin{eqnarray}\label{FinEq1}
&&\frac{c}{4\pi}g\nabla g \times \nabla \varphi = - \left[ \frac{c}{4\pi} \widetilde\Delta f +
r \sigma_{\wedge}(\nabla f^2 + g^2)^{1/2}\frac{d \Psi}{d f} \right]\cdot \\  \nonumber
&&(\nabla f \times \nabla \varphi)-\sigma_{\bot} g \frac{d \Psi}{d f}\nabla f
\end{eqnarray}
The right hand side of the equation (\ref{FinEq1}) has two components: along the vector $\nabla f$ and along the vector
$\nabla f\times\nabla\varphi$, which is orthogonal to it.
So, it is convenient to introduce the orthogonal coordinate system $(f, x_1, \varphi)$, where $\nabla x_1 = \nabla f \times \nabla \varphi$.
The coordinate $x_1$ is directed along the magnetic surface $f=const$. In virtue of the cylindrical symmetry the function
$g$ is a function of two coordinates, $f$ and $x_1$, $g(f,x_1), \, \nabla g=(\partial g/\partial f)\nabla f+(\partial g/\partial x_1)\nabla x_1$.
Then the component of the equation (\ref{FinEq1}) along $\nabla f$ gives
$$
g=\frac{4\pi}{c}\frac{d\Psi}{df}\int_0^{x_1} r^2(f,x'_1)\sigma_\perp(f,x'_1)dx'_1+F(f).
$$
We consider the coordinate $x_1$ is equal to zero at the stellar surface $h=0$. The function $F(f)$ is determined by the distribution of
the accretion electric current $j_0(r)$, $j_0\simeq I/2\pi r_0^2$,
incident upon the surface,
$$
F(f)=\frac{4\pi}{c}\int_0^f j_1(f')df',
$$
where the value of $j_1(f)$ is the current density as a function of the magnetic flux $f, \, j_1\nabla x_1 = j_0{\bf e}_h$. The total electric
current is zero, i.e. $\int_0^{f_0} j_1(f')df'=0$. The value of $f_0$ is the boundary value of the magnetic flux, which limits the region of
existence of the electric field, $f<f_0$. As a result, for the toroidal magnetic field, $B_T=g/r$, we get the following expression
\begin{equation}\label{TorB}
g=\frac{4\pi}{c}\left(\frac{d\Psi}{df}\int_0^{x_1} r^2\sigma_\perp dx'_1+\int_0^f j_1 df'\right).
\end{equation}
The first term in the equation (\ref{TorB}) is the toroidal magnetic field generated by the transverse electric current $j_\perp$. It is estimated
as follows. Emerging electric field $E$ is determined by the ratio $j_\perp=\sigma_\perp E$. On the other hand, under the continuity
current condition $j_\perp\simeq rj_0/h, \, h>r$, and $E\simeq rj_0/h\sigma_\perp$. Thus, $d\Psi/df\simeq j_0/h\sigma_\perp B_P$, where the
value of $B_P$
is the poloidal magnetic field, $B_P=|\nabla f|/r$. Due to the strong elongation of the magnetic surfaces, $h>>r$, the coordinate $x_1$
almost
coincides with the depth $h$, $dx_1\simeq B_P dh$. Therefore, $g\simeq 4\pi j_0r^2/c$, $B_T\simeq 4\pi j_0r/c$. The same toroidal field creates
the longitudinal current $j_0$ (the second term in equation (\ref{TorB})). But the poloidal magnetic field exceeds the toroidal one due to the
fact that the Hall conductivity is
significantly larger than the transverse conductivity, $\sigma_\wedge>>\sigma_\perp$.
The toroidal current is of
$j_\phi=\sigma_\wedge E\simeq \sigma_\wedge r
j_0/\sigma_\perp h$, it generates the poloidal magnetic field $B_P\simeq (4\pi/c)(\sigma_\wedge/\sigma_\perp)r^2j_0/h$,
which near by the surface
$h\simeq r_0$ is $B_{P0}\simeq (4\pi/c)(\sigma_\wedge/\sigma_\perp)r_0j_0=B_0$.
We see that the poloidal field $\sigma_\wedge/\sigma_\perp$ times larger than the toroidal field.

\begin{figure}
\begin{center}
\includegraphics[width=8cm]{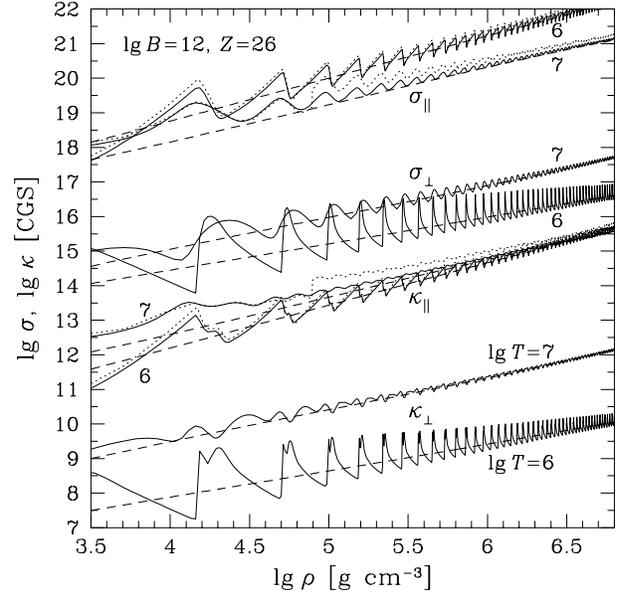}
\end{center}
\caption{Conductivities along $\sigma_{\Vert}$ and across the magnetic field $\sigma_{\bot}$
depending on the density $\rho$. The magnetic field is $B=10^{12}$ G (Potehkin 1999) }
\label{fig3}
\end{figure}

\begin{figure}
\begin{center}
\includegraphics[width=8cm]{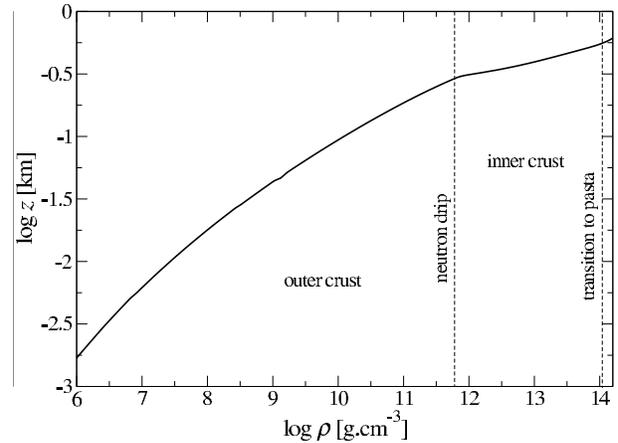}
\end{center}
\caption{Density of the crust $\rho$ versus the depth $h$ (Chamel \& Haensel 2008)}
\label{fig4}
\end{figure}

The dependence of the poloidal field on coordinates $r, \, h$ is given by the equation
\begin{equation}\label{PoloidalFromRH}
\widetilde\Delta f +g\frac{\partial g}{\partial f} + \frac{4 \pi r \sigma_{\wedge}}{c}(\nabla f^2 + g^2)^{1/2}\frac{d \Psi}{d f}=0,
\end{equation}
which is the projection of the equation (\ref{FinEq1}) onto the direction of $\nabla f\times\nabla\varphi$. The second term in this equation describes
the generation of the poloidal
field by the longitudinal electric current flowing along the total magnetic field, and thereby, partly along the toroidal magnetic field.
In view of above estimations, it is $(\sigma_\perp/\sigma_\wedge)^2$ times less than the third term in the equation (\ref{PoloidalFromRH}) that
is responsible
for the generation of the poloidal field by the Hall current. The electric field $d\Psi/df$ is a function of the poloidal magnetic flux $f$
and should be
zero at the boundary $f=f_0$, where the toroidal magnetic field is also zero (see the expression (\ref{TorB})). Because of the discussed problem
is linear with respect to the electric current and hence to the magnetic field, we have
$$
\frac{d\Psi}{d f}=\kappa(f-f_0),
$$
where $\kappa$ is a constant not depending on $f$ and coordinates $r,h$. According to above estimations
$\kappa=j_0/r_0^3\sigma_\perp B_{0}^2$. Standing in the equation (\ref{PoloidalFromRH}) the product of $r\sigma_{\wedge}(\nabla f^2 + g^2)^{1/2}$
is equal to
$r^2\sigma_\wedge B$, which does not depend on the magnetic field $B$. The Hall conductivity for magnetized electrons is equal to
$\sigma_\wedge=\sigma_\parallel/(\omega_c\tau_e)$, where $\omega_c$ is the cyclotron frequency of electrons and $\tau_e$ is equal to their
relaxation time.
But the conductivity increases with the depth $h$, i.e. with the increase of the density of the matter $\rho$. For example, the dependence
of longitudinal and transverse conductivities,
as well as the thermal conductivities for two values of temperatures ($T=10^6$ K and $T=10^7$ K), on the density is shown in Fig. \ref{fig3}
(Potehkin 1999). The similar behavior have these values at $T=10^8$ K. Averaging over quantum oscillations gives the linear dependence of
the Hall conductivity on the density (Potehkin 1999)
\begin{eqnarray}
\log{\sigma_{\wedge}}=1.000\log{\rho}+12.62 \, (T=10^6 K), \\   \nonumber
\log{\sigma_{\wedge}}=1.015\log{\rho}+12.56 \, (T=10^7 K), \\  \nonumber
\log{\sigma_{\wedge}}=0.983\log{\rho}+10.75 \, (T=10^8 K).
\end{eqnarray}
Thus, $\sigma_\wedge B={\sigma_\wedge}_0 B_{0}(\rho/\rho_0)$. Here the index '0' indicates the values on the surface of the
star ($h=0$). Let us introduce the dimensionless distances $r'$ and $h'$,  $r'=r/r_0, \, h'=h/r_0$. As a result, the equation (\ref{PoloidalFromRH})
becomes equal to
\begin{equation}\label{Dimless}
\widetilde\Delta'(f-f_0) + r'^2\frac{\rho}{\rho_0}(f-f_0)=0.
\end{equation}
The density $\rho/\rho_0$, standing in the equation (\ref{Dimless}), is a function of the depth $h$ and can be approximated by a power-law
dependence (Chamel \& Haensel 2008)
\begin{equation}
\rho=\rho_0\left[1+\left(\frac{h}{L}\right)^{2.5}\right],
\end{equation}
where $L$ is the characteristic scale of change of the density of the crust, $L\simeq 2.1$ m, $\rho_0\simeq 10^4 g/cm^3$. The dependence
of the density of the crust on the depth is shown in Fig. \ref{fig4}.

At large depths, $h>r_0$, the derivative over $h$ in the equation (\ref{Dimless}) becomes small compared with the derivative over $r$.
Therefore, the equation (\ref{Dimless}) is simplified
\begin{equation}\label{SimpleDimless}
r'\frac{\partial}{\partial r'}\left(\frac{1}{r'}\frac{\partial}{\partial r'}\right)(f-f_0)+r'^2\frac{\rho}{\rho_0}(f-f_0)=0.
\end{equation}
Introducing the variable $z=r'^2(\rho/\rho_0)^{1/2}/2$, the solution of the equation (\ref{SimpleDimless}) is easily obtained
\begin{equation}
f-f_0=f_0(a \, sinz+b \, cosz),
\end{equation}
where $a$ and $b$ are dimensionless constants that are determined from the boundary conditions. On the axis $r=0 \, (z=0)$ the value of the flux is
zero, $f=0$.
Then $b=-1$. Near the surface, $\rho\simeq\rho_0$, the magnetic field should be close to uniform, $f=B_0r^2/2$. This condition determines
the constant $a$, $a\simeq 2$. Then the boundary $f=f_0$ is determined by the condition $tan z\simeq 1/2, \, z\simeq 0.46$, which with
a good accuracy
gives the following dependence of the thickness of the 'magnetic needle' and the magnitude of magnetic field inside it on the depth,
\begin{equation} \label{BfromH}
r=r_0\left(\frac{\rho}{\rho_0}\right)^{-1/4}, \, B=B_0\left(\frac{\rho}{\rho_0}\right)^{1/2}.
\end{equation}
The obtained dependences have a simple explanation. The poloidal
magnetic field $B_P\simeq B$ is generated by the Hall current,
$B/r=4\pi\sigma_\wedge\Psi/rc$. On the other hand, on the  stellar
surface $B_0=4\pi\sigma_{\wedge 0}\Psi/c$. Thus,
$B=B_0(\sigma_\wedge/\sigma_{\wedge 0})$. But $\sigma_\wedge
B/\rho=\sigma_{\wedge 0} B_0/\rho_0$, and we obtain
$B=B_0(\rho/\rho_0)^{1/2}$. Because the flux of the magnetic field
is conserved, $r=r_0(\rho/\rho_0)^{-1/4}$.

\begin{figure}
\begin{center}
\includegraphics[width=8cm]{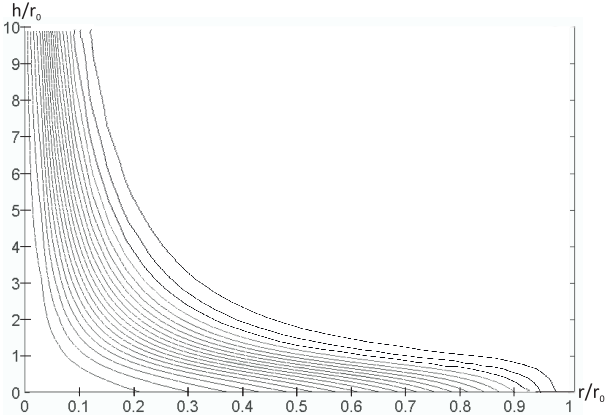}
\end{center}
\caption{Lines of the constant poloidal magnetic flux, $f(r',h')=const, f(r',h')$ is the solution of the equation (\ref{Dimless}). Values $r'$ and $h'$ are dimensionless coordinates $r'=r/r_0, \, h'=h/r_0$.
The function $f(r',h'=0)=f_0 r'^2$ describes the uniform vertical magnetic field $B_0=2f_0/r_0^2$ at the stellar surface in the polar region.}
\textbf{\label{fig5}}
\end{figure}

The numerical solution of the equation (\ref{Dimless}) under the
boundary condition $f(h=0)=B_0r^2/2$ confirms these dependence
(Fig. \ref{fig5}). We see that the accreting electric current
really penetrates into the inner crust and the core of a star, while greatly narrowing
and amplifying the magnetic field. When the density becomes
$\rho\simeq 10^{14} g cm^{-3}$ the size of the magnetic tube
decreases more than two orders of magnitude ($r<r_0/10^2\simeq 10
cm$), and the magnetic field is enhanced by five orders of
magnitude ($B\simeq 10^5 B_0$). Thus, the magnetic field in the
core in a small area can reach values $10^{15} - 10^{17}$ Gauss that essentially affects on
properties of the core matter.

\section{Structure of the crust and the core of a neutron star}

In order to understand how the super strong magnetic field
$B>10^{15}$ Gauss has an influence upon the properties of the
stellar matter we have to know the structure of a neutron star.

According to modern models, a neutron star has a solid crust and a
liquid superfluid and super conducting core (Shapiro \& Tuekolsky 1983; Chamel \& Haensel 2008).
The crust of a neutron star consists of inner ($Aen$-phase) and
external ($Ae$-phase) parts. $Ae$-phase consists mainly of iron
nucleus $Fe_{56}$ and degenerate gas of free electrons. Because of
the mutual electrostatic repulsion, nucleus of iron form the
body-centered crystal lattice, thereby creating a hard outer crust
of neutron stars. The density of the matter in $Ae$-phase changes
from $10^4$ $\text{g/cm}^3$ to $4.3\cdot10^{11}$ $\text{g/cm}^3$.
$Aen$-phase contains neutron-rich nuclei, which form another
lattice, degenerate gases of free relativistic electrons and free
neutrons. The density of the matter in $Aen$-phase
changes from $4.3\cdot10^{11}$ to $2.4\cdot10^{14} \text{g/cm}^3$.
The total thickness of the crust is about one kilometer
(Shapiro \& Tuekolsky, 1983). The important property of the crust and also the core
is the absence of the electron superconductivity at real stellar
temperatures even at high pressures of the stellar matter (Chamel \& Haensel 2008).

When the density of the matter becomes about nuclear one, nuclei
are destroyed and form the so-called $npe$-phase, which is a
homogeneous mixture of neutron, proton and electron fluids.
The density of protons and electrons are equal in the
frame where the star is at rest, and make up a few percent of the
density of neutrons (Chamel \& Haensel 2008). In the inner crust, at the
modern view, there is the superfluid matter. This is mainly
neutrons and protons. Superfluidity of charged protons is
associated with superconductivity. Superconductivity of protons in
the core of a neutron star is described by the theory of
Ginzburg-Landau (1950). Comments about existence of super conducting
vortices in a neutron star core were made by Ginzburg and
Kirzhnits (1965). Typically, the proton coherence length
in the stellar core, which is of 50 fm, smaller than the
penetration length, which is of 100-300 fm. Thus, the system of
superconductors of the second kind is formed in the core
(Yakovlev, 2001). The magnetic flux permeates the matter of a
neutron star through producing regions in which superconductivity
is suppressed. The structure of these regions depends on the type
of a superconductor. The type of superconductivity depends on the
ratio of two scales: 1) the coherence length of the proton
$\xi_p$, at which Cooper's pairs of protons can be destroyed by
quantum fluctuations, and 2) the London's penetration depth
$\Lambda_l$. If $\sqrt{2}\Lambda_l>\xi_p$ then the
superconductivity must be of the second kind. Then, it is
favorable to form a set of super conducting vortices of radius
$\sim \xi_p$ with a core of usual protons. Each vortex carries one
quantum of magnetic flux $\phi_0=hc/2e\simeq 2\cdot10^{-7}G cm^2$.
For neutron stars values of $\xi_p$ and $\Lambda_l$ are equal to
$$
\xi_p\approx 5\cdot 10^{-12}\left(\frac{m^{*}_p}{m_p}\right)^{-1}\rho_{14}^{1/3}\left(\frac{x_p}{0.1}\right)^{1/3}
\left(\frac{T_{cp}}{10^9 K}\right)^{-1} \, cm;
$$
$$
\Lambda_l \approx 9\cdot 10^{-12}\left(\frac{m^{*}_p}{m_p}\right)^{1/2}
\rho_{14}^{-1/2}\left(\frac{x_p}{0.1}\right)^{-1/2} \, cm,
$$
where $\rho_{14}=\rho/10^{14}g\cdot cm^{-3}$ is the density of the matter in units of $10^{14} g cm^{-3}$, $x_p$ is the fraction of protons,
$m^{*}_p$ is the effective
proton mass, $m^{*}_p\simeq 0.6m_p$, $T_{cp}$ is the critical temperature for superconductivity of protons.
Substituting the values typical for neutron stars, we obtain the value of the second critical field $B_{c2}$,
\begin{equation}\label{CriticalField}
B_{c2}=\frac{\phi_0}{2\pi\xi_p^{2}}\approx 10^{15}\left(\frac{m^{*}_p}{m_p}\right)\left(\frac{x_p}{0.1}\right)^{-2/3}\rho_{14}^{-2/3}
\left(\frac{T_{cp}}{10^9 K}\right)^2 \, G,
\end{equation}
which suppresses the proton superconductivity. In this field the
normal cores of vortices begin to touch each other that leads to
the complete disappearance of the superconductivity (Glampedakis,
Andersson \& Samuelsson 2011).

\section{Motion of super conducting vortices}

\begin{figure}
\begin{center}
\includegraphics[width=8cm]{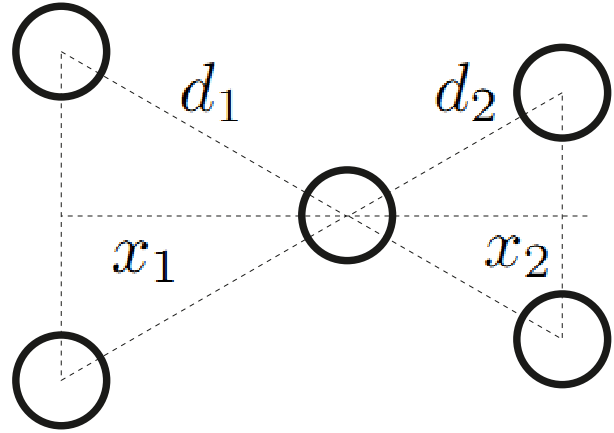}
\end{center}
\caption{Arrangement of magnetic vortices}
\label{fig6}
\end{figure}

We saw that the super strong magnetic field generated by the
accretion current abolishes super conducting vortices in the inner
crust and the core. But it  happens only inside the narrow region
of the radius of 10 centimeters in the core near the magnetic
dipole axis. However, this small volume contains macroscopic
numbers of magnetic vortices, more than $10^{21}$. And their lack
must influence upon other vortices in the core.

As we wrote above, the superconductivity of the second kind takes place in the inner
crust and the core of neutron stars (Baym, Pethick \& Pines 1969). Magnetic
vortices are arranged in the triangular lattice (Shapiro \& Tuekolsky 1983). Let us consider the interaction of three vortices located
at apexes of the equilateral triangle.
The repulsive force per unit length, acting on the third vortex from the first two, is equal to
$$
F_l=\frac{\phi_0^2}{4\pi^2\Lambda_l^2}\frac{x}{x^2+(d/2)^2}
$$
Here $d$ is the distance between two neighbour vortices, $x$ is
the distance to third vortex (see Fig. \ref{fig6}). In the case of
uniform distribution of vortices the total repulsive force is
equal to zero. However, near the empty region the repulsive force is not
compensated, and vortices must move into the region where their density
is zero, i.e. inside the dead zone. So, the gradient of the vortices concentration
appears in the volume outside of the region of the 'magnetic needle'.
The density of vortices $n$ is determined by the strength of the magnetic field,
$n=B_0/\phi_0$. So, for the magnetic field $B_0=10^{12} G$ the
density of vortices is equal to $n=5\cdot 10^{18} cm^{-2}$. The
force per unit length, acting on the vortex from neighbours in the
presence of the gradient of concentration, is equal to (Schmidt 1997)
\begin{equation}\label{FlucForce}
F_l=\frac{\phi_0^2}{4\pi^2\Lambda_l^2}\left(\frac{x_1}{x_1^2+\left(d_1/2\right)^2}-\frac{x_2}{x_2^2+\left(d_2/2\right)^2}\right).
\end{equation}
From the geometry one can see that $x_1=\sqrt 3 d_1/2$ and $x_2=\sqrt 3 d_2/2$. Substituting this values into the equation (\ref{FlucForce})
we obtain
$$
F_l=\frac{\sqrt 3}{2}\frac{\phi_0^2}{4\pi^2\Lambda_l^2}\left(\frac{1}{d_1}-\frac{1}{d_2}\right).
$$
Considering that the difference $d_2-d_1$ is small compared with $d$, in the first approximation we obtain
$$
F_l=\frac{\sqrt 3}{2}\frac{\phi_0^2}{4\pi^2\Lambda_l^2}\left(\frac{d_2-d_1}{d^2}\right),
$$
Let us find the dependence of $d$ on the concentration $n$. The area occupied by the triangle with three vortexes placed in apexes
is $S=xd/2=\sqrt 3 d^2/4$.
Thus, $n=1/2S=2/(\sqrt 3 d^2)$. Let us find the dependence of $d_2-d_1$ on $\nabla n$, $\nabla n=\Delta n/d=-4(d_2-d_1)/(\sqrt 3 d^4)$.
As a result, the force $F_l$ is equal to
\begin{equation}
{\bf F}_l=-A\frac{\nabla n}{n},
\end{equation}
where $A$ is
\begin{equation}
A=\frac{\sqrt 3}{16\pi^2}\frac{\phi_0^2}{\Lambda_l^2}=4.4\cdot 10^6\left(\frac{\Lambda_l}{10^{-11} cm}\right)^{-2} \, din.
\end{equation}
Vortices are moving in the electron gas, which is not super
conducting at the core temperature. If the velocity of vortices is
not equal to zero, the electron gas creates the significant frictional
force. This force per unit length, $F_f$, is equal to
(Alpar, Langer \& Sauls 1984)
$$
{\bf F}_f=-\frac{\rho_e}{n \tau_e}({\bf v}-{\bf v}_e),
$$
where $\tau_{e}=10^{-14}$ s is the electron relaxation time, $\rho_e=(m_e/m_p)x_p\rho=
5.4\cdot 10^9\rho_{14}(x_p/0.1) g  cm^{-3}$ is the density of electrons in the core of stars, ${\bf v}$ is the velocity of vortices.
Considering the electron liquid is stationary relative to the star, ${\bf v}_e=0$, we get
\begin{equation}
{\bf F}_f=-\nu{\bf v}, \,\nu=\frac{\rho_e}{n\tau_e}.
\end{equation}
Of the order of magnitude the value of $\nu$ is equal to
\begin{equation}
\nu=1.1\cdot 10^5\rho_{14}\left(\frac{x_p}{0.1}\right)\left(\frac{n}{5\cdot 10^{18}cm^{-2}}\right)^{-1} \, g \, cm^{-1} s^{-1}.
\end{equation}
Let us now consider the motion of super conducting vortices in the
presence of the region in which the magnetic field exceeds the
second critical value. This narrow area of the size
$r^*=r_0(\rho/\rho_0)^{-1/4}<<r_0<<R$ \label{rstar} is near by the
axis of the dipole. In this area the superconductivity breaks
down, and the density of vortices becomes zero. Thus, outside,
$r>r^*$, there is a gradient of vortex density, and vortices are
moving to the center. In virtue of the cylindrical symmetry the
velocity is directed along the radius, $v_r\equiv v$. The vortex
motion is described by equations
\begin{eqnarray}\label{VortexDyn}
&&m\left(\frac{\partial v}{\partial t}+v\frac{\partial v}{\partial r}\right)=-\nu v-\frac{A}{n}\frac{\partial n}{\partial r}, \\
&&\frac{\partial n}{\partial t}+\frac{1}{r}\frac{\partial}{\partial r}\left(rvn\right)=0.
\end{eqnarray}
Here $m$ is the mass of a vortex per unit length (Glampedakis, Andersson \& Samuelsson 2011),
$$
m=7\cdot 10^{-10}\left(\frac{m_p^*}{m_p}\right)^{-2}\rho_{14}^{5/3}\left(\frac{x_p}{0.1}\right)^{5/3}\left(\frac{T_{cp}}{10^9 K}\right)^{-2} \, g/cm.
$$
The maximum speed of vortices at $r\simeq r^*$ is of the order of $|v|\simeq A/\nu r^*\simeq 4 cm/s$.
Correspondingly, the acceleration is
equal to $v^2/r^* \simeq 2cm/s^2$. Thus, the inertia of vortices could be neglected. Their speed is determined by the balance between the
pressure and the friction force
$$
v=-\frac{A\tau_e}{\rho_e}\frac{\partial n}{\partial r}.
$$
The vortex flux $J$ flowing into the region $r<r^*$ is $J=2\pi rvn$,
\begin{equation}
J=-2\pi\frac{A\tau_e}{\rho_e}rn\frac{\partial n}{\partial r}\simeq -\pi\frac{A\tau_e}{\rho_e}n^2.
\end{equation}
Vortices are destroying in the area $r<r^*$, their number is constantly decreasing in time
\begin{equation}\label{VortexEv}
\pi R^2\frac{d n}{dt}=-\pi\frac{A\tau_e}{\rho_e} n^2.
\end{equation}
The solution of equation (\ref{VortexEv}) gives the exponential vortex density decrease
\begin{equation}\label{SolvDen}
n=n_0\left(1+\frac{t}{\tau}\right)^{-1}, \, \tau=\frac{\rho_e R^2}{A\tau_e n_0}.
\end{equation}
Here $n_0$ is the initial density at $t=0$. The time $\tau$, as we will see below, the formula (\ref{FinalB}),
is the large time of the order of $10^3$ years and is explained by the small scale of the dead zone $r*\simeq 10$ cm.
Indeed, the speed of vortex motion near the magnetic axis is of 4 cm/s, and the time to overpass the distance of the stellar
radius $R\simeq 10^6$ cm is of $2\cdot 10^5$ s. But the area of the 'magnetic needle' $2\pi r*$ is much
smaller than $2\pi R$, $r*/R\simeq 10^{-5}$,
and we just obtain the characteristic time of the magnetic vortex density and the magnetic field evolution of $2\cdot 10^{10}$ s.

\section{Evolution of magnetic fields}

Since the magnetic field of the neutron star $B_0$ and the density
of super conducting vortices in the core $n$ are related by the
relation $B_0=n\phi_0$, then the evolution of the magnetic field
strength occurs at the same rate of the vortex density changing
(\ref{SolvDen})
\begin{equation}
B_0=B_0^i\left(1+\frac{t}{\tau}\right)^{-1}, \, \tau=\frac{\rho_e \phi_0 R^2}{A\tau_e B_0^i}.
\end{equation}
The typical magnetic field evolution time $\tau$ is equal to
\begin{equation}\label{FinalB}
\tau=\frac{16\pi^2}{3^{1/2}}\frac{\rho_e\Lambda_l^2 R^2}{\tau_e\phi_0 B_0^i}=2\cdot 10^{10}\left(\frac{m_p^*}{m_p}\right)\left(\frac{R}{10^6 cm}\right)^2\left(\frac{B_0^i}{10^{12} G}\right)^{-1} \, s.
\end{equation}
The most surprising result is the time $\tau$ (\ref{FinalB}), determined by microscopic scales of the proton superconductivity
($\Lambda_l\simeq 10^{-11} cm, \, \phi_0\simeq 10^{-7} G cm^2$), has the reasonable astrophysical value.

We see that the magnetic field half-life is $10^3$ years for the initial field about
$10^{12}$ Gauss. However, this time increases for weaker initial fields $B_0^i$. One can see that the time $t_a$
to achieve the field $B_0$, does not depend on the initial field ($B_0^i>B_0$). This time is determined by the relation
$t_a=\tau(B_0^i=B_0)$,
$$
B_0=0.6\cdot 10^{12}\left(\frac{m_p^*}{m_p}\right)\left(\frac{R}{10^6 cm}\right)^2\left(\frac{t_a}{10^3 y}\right)^{-1} G.
$$
So, it takes $10^6$ years of accretion onto a neutron star to
achieve the magnetic field $10^9 G$. Thus, the magnetic field of
neutron stars that have passed an accretion phase is determined
basically by the accretion time. However, it should be noted, that
to reduce the magnetic field of the star, the magnetic field at
the top of the 'magnetic needle' must exceed the second critical
field $B_{c2}$ (\ref{CriticalField}) for the proton
superconductivity destruction. This means that there is a lower
limit for the final magnetic field $B_0^{min}$, which is
determined by the condition
$B_0^{min}=B_{c2}(\rho_0/\rho_c)^{1/2}$. Here $\rho_c$ is the core
matter density, in which super conducting vortices of the second
kind exist.

At the end of this section we have to make one more remark: the
elementary magnetic flux of the vortex $\phi_0$, does not
disappear immediately after its destruction. It is stored in the
form of a magnetic field generated by an electric current flowing
at the boundary of the super conducting region $r=r^*$. This
current is not super conducting. Therefore, it fades over the time
$4\pi\sigma (r^*)^2/c^2$, where $\sigma$ is the electron
conductivity of the core. Since the current flows across the
magnetic field $\sigma=\sigma_\perp$. Estimating the transverse
conductivity as $\sigma_\perp\simeq 10^{22}(B_0/10^{12}G)^{-2} s$,
we find that the current decay time, $10^4(r^*/10 cm)^2
(B_0/10^{12}G)^{-2} s$, is much less than the magnetic field
evolution time $\tau$. So, the disappearance of the vortex in the
time scale of $10^3$ years could be considered as instantaneous.

\section{Conclusion}

We proposed a new mechanism for a neutron star magnetic field
decay by accretion of matter onto a star. The mechanism of field decay
is the accretion of a matter, which is accompanied by an electric
current, generated by the rotating magnetic field in the system
star - accretion disk. Closed electric current loop is supported
by the voltage, which is generated by the rotation of the stellar
magnetic field. This current transfers the angular momentum from
the disk to the star, spinning it up. To transfer the angular
momentum, the current have to flow through the stellar surface.
But the conductivity of the crust along the magnetic field is
significantly higher than the conductivity across the field, along
which the current must be closed. Thus, the electric current
penetrates deep into the star. In addition, the Hall conductivity
is also significantly higher than the transverse conductivity -
therefore, the electric current in the crust flows mainly in the
azimuthal direction. So, the longitudinal magnetic field greatly
increases. In the crust polar cap there occurs the narrow area, in
which the magnetic field increases with depth,
$B=B_0(\rho/\rho_0)^{1/2}$. The area of the electric current
closing, in which a strong magnetic field  is inside the narrowing
magnetic tube (the 'magnetic needle' - see Fig. \ref{fig2}), has
the radius decreasing with increasing of the conductivity, and
hence with increasing of the crust density,
$r=r_0(\rho/\rho_0)^{-1/4}$. If magnetic field at the end of the
needle exceeds the second critical field $B_{c2}$, which destroys
the second kind superconductivity of protons, then magnetic
vortices inside the tube $r<r^*$ (\ref{BfromH}) disappear. A vortex
density gradient appears, it leads to the vortex motion towards
their death area. Thus, the density of vortices in the inner crust
and in the core continuously decreases in time. This process
reduces the magnetic field of the star $B_0$. The decay law has a
power law $B_0=B_0^i(1+t/\tau)^{-1}$ with the characteristic time
of $\tau\simeq 10^3 (B_0^i/10^{12} G)^{-1}$ years , which depends on
the initial magnetic field $B_0^i$. As a result, the final
magnetic field after accretion of matter (plasma), depends mostly
on the time of accretion $t_a$ and is inversely proportional to it
(\ref{FinalB}). Here it should be emphasized that in a wide range
of accretion rate ${\dot M}$ the evolution of the magnetic field
does not depend on the accretion rate. The point is that the
accretion electric current $I$ (\ref{Current}) is significantly
less than the maximum current $e{\dot M}/m_p$. There are more than
six orders reserve of the magnitude of the accretion rate.
However, the magnetic field of the star can't be reduced to small
values, while accretion can last a long time. If the field in the
inner part of the 'magnetic needle' becomes less than the critical
field, then the destruction of the proton superconductivity does
not occur. The magnetic field stops to decay. This effect was
actually observed. There is no recycled stars with magnetic fields
less than $B_0^{min}\simeq 10^8 G$ (see Fig. \ref{fig1}). Using
the expression for critical field (\ref{CriticalField}) and the
expression for the magnetic field in the 'magnetic needle'
(\ref{BfromH}), we obtain the following estimation
\begin{eqnarray}
&&\left(\frac{m^{*}_p}{m_p}\right)\left(\frac{x_p}{0.1}\right)^{-2/3}\rho_{14}^{-2/3}\left(\frac{10^{10}\rho_0}{\rho_c}\right)^{1/2}
\cdot \\  \nonumber
&&\left(\frac{T_{cp}}{10^9 K}\right)^2\left(\frac{B_0^{min}}{10^8 G}\right)^{-1} \simeq 10^{-2}, \nonumber
\end{eqnarray}
which is not critical. For example, if $\rho_{14}\simeq 10$ in the center of the star and $T_{cp}\simeq 3\cdot 10^8$ K the proposed model
does not contradict the observed value $B_0^{min}\simeq 10^8 G$. It should to be noted that to achieve the magnetic field $B_0\simeq 10^8 G$
it is necessary that the accretion continues at least $t_a\simeq 6\cdot 10^6$ years.

We see that in the proposed model of attenuation of the magnetic field of accreting neutron stars the decisive role plays the
internal  structure of neutron stars. A comparison of observational data of accreting or passed a stage of accretion neutron stars
with the proposed model could give information about the physical parameters of the structure of the neutron star. But it is beyond the scope of
this paper.

\section*{Aknowlegement}

This work was done under support of the Russian Foundation for Fundamental
Research (grant numbers 13-02-12103 and 15-02-03063).


\begin{thebibliography}{99}

\bibitem{alpar}
Alpar M. A., Langer S. A. Sauls, J. A., 1984, ApJ, 282, 533
\bibitem {pines}
Baym G., Pethick C., Pines D., 1969, Nature, 224, 5220
\bibitem {beskin-gurevich}
Beskin V.S., Gurevich A.V., Istomin Ya.N., 1993, Physics of the Pulsar Magnetosphere, Cambridge University Press
\bibitem {istomin-philipov}
Beskin V.S., Istomin Ya.N., Philippov, A.A., 2013, Phys. Uspekhi, 56, 164
\bibitem {Bland}
Blandford R.D., Applegate J.H., Hernquist L., 1983, MNRAS, 204, 1025
\bibitem{hansel}
Chamel N., Haensel P. 2008, Living Reviews in Relativity, 11, N 10, 59
\bibitem {GL}
Ginzburg V.L,, Landau L.D., 1950, Zh. Eksp. Teor. Fiz., 20, 1064
\bibitem {kirzhnic}
Ginzburg, V.L., Kirzhnits, D.A., 1965, JETP, 20, 1346
\bibitem {glampedakis}
Glampedakis K., Andersson N., Samuelsson L., 2011, MNRAS, 410, 805
\bibitem {goldreich}
Goldreich P., Julian W.H., 1969, ApJ, 157, 869
\bibitem {hewish}
Hewish A., Bell S.J., Pilkington J.D.H., Scott P.F., Collins, R.A., 1968, Nature, 217, 709
\bibitem {istomin}
Istomin Ya.N., Smirnov A.P., Pak D.A., 2005, MNRAS, 356, 1149
\bibitem {istomin-haensel}
Istomin Ya.N., Haensel, P., 2013, Astronomy Rep., 57, 904
\bibitem{Landau} 
Landau, L.D., Lifshitz, E.M., Pitaevskii, L.P., 1984, Electrodynamics of continuous media,
Elsevier, Butterworth Heinemann, 220
\bibitem {lorimer}
Lorimer D.R., 2001, Living Reviews in Relativity, 4, N 5, 12
\bibitem {Lyne-Anderson}
Lyne A.G., Anderson B. Salter M.J., 1982, MNRAS, 201, 503
\bibitem {potekhin}
Potekhin A.Y., A\&A, 351, 787
\bibitem {shmidt}
Schmidt V.V., 1997, The Physics of Superconductors, Springer-Verlag, Berlin, Heidelberg
\bibitem {shapiro}
Shapiro S.L., Tuekolsky S.A., 1983, Black Holes, White Dwarfs and Neutron Stars, John Wiley and Sons
\bibitem {yakovlev}
Yakovlev, D.G., 2001, Phys. Uspekhi, 44, 823

\end{thebibliography}
\end{document}